\documentclass[twocolumn,english,prb,superscriptaddress,showpacs]{revtex4}
\usepackage[T1]{fontenc}
\usepackage[latin9]{inputenc}
\usepackage{amssymb}
\usepackage{graphicx}

\makeatletter
\@ifundefined{textcolor}{}
{%
 \definecolor{BLACK}{gray}{0}
 \definecolor{WHITE}{gray}{1}
 \definecolor{RED}{rgb}{1,0,0}
 \definecolor{GREEN}{rgb}{0,1,0}
 \definecolor{BLUE}{rgb}{0,0,1}
 \definecolor{CYAN}{cmyk}{1,0,0,0}
 \definecolor{MAGENTA}{cmyk}{0,1,0,0}
 \definecolor{YELLOW}{cmyk}{0,0,1,0}
}

\makeatother

\usepackage{babel}
\begin{document}

\title{Magnetoelastic effects in doped Fe$_{2}$P}

\author{Z. Gercsi}

\affiliation{Dept. of Physics, Blackett Laboratory, Imperial College London, London
SW7 2AZ, United Kingdom}

\author{E. K. Delczeg-Czirjak}

\affiliation{Division of Materials Theory, Department of Physics and Astronomy,
Uppsala University, Box 516, SE 75120 Uppsala, Sweden}

\author{L. Vitos}

\affiliation{Applied Materials Physics, Department of Materials Science and Engineering,
Royal Institute of Technology, SE-100 44 Stockholm, Sweden}

\author{A. S. Wills}

\affiliation{Department of Chemistry, University College London, 20 Gordon Street,
London WC1H 0AJ, United Kingdom}

\author{A. Daoud-Aladine}

\affiliation{ISIS facility, Rutherford Appleton Laboratory, Chilton, Didcot, Oxfordshire,
OX11 0QX, United Kingdom}

\author{K.G. Sandeman}

\affiliation{Dept. of Physics, Blackett Laboratory, Imperial College London, London
SW7 2AZ, United Kingdom}

\begin{abstract}
We use combine high resolution neutron diffraction (HRPD) with density
functional theory (DFT) to investigate the exchange striction  at
the Curie temperature ($T_{C}$) of Fe$_{2}$P and to examine the
effect of boron and carbon doping on the P site. We find a significant
contraction of the basal plane on heating through $T_{C}$ with a
simultaneous increase of the $c$-axis that results in a small overall
volume change of $\sim0.01\%$. At the magnetic transition the Fe$_{I}$-Fe$_{I}$
distance drops significantly and becomes shorter than Fe$_{I}$-Fe$_{II}$.
The shortest metal-metalloid (Fe$_{I}$-P$_{I}$) distance also decreases
sharply. Our DFT model reveals the importance of the latter as this
structural change causes a redistribution of the Fe$_{I}$ moment
along the $c$-axis (Fe-P chain). We are able to understand the site
preference of the dopants, the effect of which can be linked to the
increased moment on the Fe$_{I}$-site, brought about by strong magneto-elasticity
and changes in the electronic band structure. 
\end{abstract}
\maketitle

\section{Introduction}

Fe$_{2}$P-based magnetic alloys attract interest from many areas
of physics, materials science and geophysics research. They are found
in meteorites and are considered as a candidate minor phase present
in the Earth's core.\cite{Fe2P_Meteorite,Fe2P_Meteor2} An understanding
of the mechanism of formation of these minerals can help to identify
the histories of planetary bodies and the composition of the Earth's
outer core. In LiFePO$_{4}$-based battery materials, a percolating
nano-network of metal-rich phosphides including Fe$_{2}$P can significantly
enhance electrical conductivity.\cite{Li-ionNature} Fe$_{2}$P-based
alloys can furthermore be prepared as 1-dimensional  nanowires and
nanocables.\cite{Fe2P_nanowire} Of most relevance to this article,
however, is the prospect of hexagonal Fe$_{2}$P-based alloys being
used as room temperature magnetic refrigerants.

Fe$_{2}$P exhibits a first order magnetic transition from a ferromagnetic
(FM) state to a paramagnetic one (PM) at 217~K~ accompanied by a
significant change in the $c$/$a$-ratio of the hexagonal structure.
The  paramagnetic susceptibility deviates from the Curie-Weiss law
for temperatures well above $T_{C}$ (up to 700~K).\cite{Fujii}
In nanocable form, the magnetic transition temperature is shifted
10-50~K higher compared to the parent composition because of strong
strain and/or carbon doping.\cite{Fe2P_nanowire} The increase in
the  $T_{C}$ with only a small, partial replacement of phosphorus
with other $p$-block elements (B, Si or As) is remarkable. 10\% replacement
of P by B leads to $\sim$120\% change, while the same amount of Si
and As substitution also results in a $\sim$70\% and $\sim$60\%
increase (Fig.~\ref{fig:doping,-Tc}) respectively, with a simultaneous
change in the nature of the transition from first order to second
order.\cite{Fe2Pdoped1,Fe2Pdoped2,Fe2Pdoped3} Such large changes
in $T_{C}$ are not restricted to doping by $p$-block elements. Partial
replacement of Fe by Mn results in a significant increase of the saturation
magnetisation, while the first order nature of the metamagnetic transition
is preserved up to and beyond room temperature and is tuned by magnetic
field at a rate of $\sim$3~KT$^{-1}$ .\cite{Dung1,Dung2}

The metamagnetism of Fe$_{2}$P shares many features with the itinerant
electron metamagnetism  of La(Fe,Si)$_{13}$.\cite{Kuzmin} That material
also has a PM to FM transition, the temperature of which can be tuned
from around 190~K to well above 300~K. The high magnetisation state
can be induced by a magnetic field above $T_{C}$, and the metamagnetic
transition is tuned by field at a rate of around 4~KT$^{-1}$. As
in Fe$_{2}$P the first or second order nature of its metamagnetism
depends on the substituent (e.g. Co or Mn) or intercolating atom (e.g.
H or C).\cite{Fujita,Fujita_H_DFT} Fe$_{2}$P-based and La(Fe,Si)$_{13}$-based
alloys are two of the leading contenders for scale-up as magnetic
refrigerants, due to their tunable metamagnetism, their large room
temperature magnetocaloric effect (MCE) and the fact that they are
mostly composed of abundantly available 3$d$ and $p$-block elements.
\cite{Karl_scripta}

There has been a recent growth in theoretical investigations of the
origin and tuneability of metamagnetim in Fe$_{2}$P and of the appearance
of a body centered orthorhombic structure in substituted alloys. This
is partly motivated by the sensitivity to doping of the $T_{C}$ and
any associated thermomagnetic hysteresis, and given added impetus
more recently by the investigation of the MCE of industrially-scaled
quantities of material where good compositional tolerance is required.
A Landau-Ginzburg free energy analysis based on fixed-spin-moment
 calculations that took into account the effect of spin fluctuations
revealed the metamagnetic nature of the Fe-atoms at a particular crystallographic
(3$f$) site in the parent alloy as well as in the doped counterparts.\cite{Yamada&Terao,Erna_Landau}
Using linear muffin tin orbitals in the atomic sphere approximation
, Severin et al. \cite{Fe2PSi_theory1} found that the magnetic moments
in the inter-related hexagonal and orthorhombic structures are very
similar and scale with the nearest-neighbour Fe-P distances. We recently
found that the phonon vibrational free energy stabilizes the hexagonal
phase, whereas the electronic and magnetic entropies favour a low
symmetry orthorhombic structure in Si-substituted Fe$_{2}$P$_{1-x}$Si$_{x}$
alloys \cite{Erna_Fe2PSi}. An analysis of the exchange constants
in the hexagonal structure of B-, Si- and As-doped Fe$_{2}$P helps
to explain trends in Curie temperature. A principal interlayer Fe-Fe
interaction was identified as controlling the strength of ferromagnetism.\cite{Erna_exchange}
Recently, Liu et al. also linked the metamagnetic transition of the
Fe moment at the 3f site to the critical Fe-Fe distances using exchange
coupling analysis based on DFT.\cite{LiuFe2PJx}

In the present paper we employ a joint experimental-theoretical approach
that we have used elsewhere to uncover the microscopic mechanism of
metamagnetism in Mn-based antiferromagnets. By using DFT in combination
with structural information from HRPD~\cite{Alex_PRL} we have previously
mapped the magnetic phase diagram of Mn-based metallic orthorhombic
compounds in the \emph{Pnma} space group as a function of Mn-Mn distance~\cite{Zsolt1}
and have predicted new antiferromagnetic metamagnets.\cite{Zsolt2,Staunton_DLM}
HRPD indicated the most relevant changes in the interatomic distances
which was then mimicked by ab-initio simulations to compare the total
energies of the competing magnetic states. Taking a similar  approach
here, we first analyse and compare the peculiar magnetoelastic coupling
of the metamagnetic transition in Fe$_{2}$P, in boron-doped Fe$_{2}$(P,B)
and, for the first time, carbon-doped Fe$_{2}$(P,C) using HRPD. In
the second part of the article, we use a simple DFT model to further
interpret the large interatomic changes at the magnetic transition.
Our aim is to provide clues as to how to control and optimise the
structure-magnetism relationship in this highly tuneable materials
class.

\begin{figure}
\includegraphics[width=8.5cm]{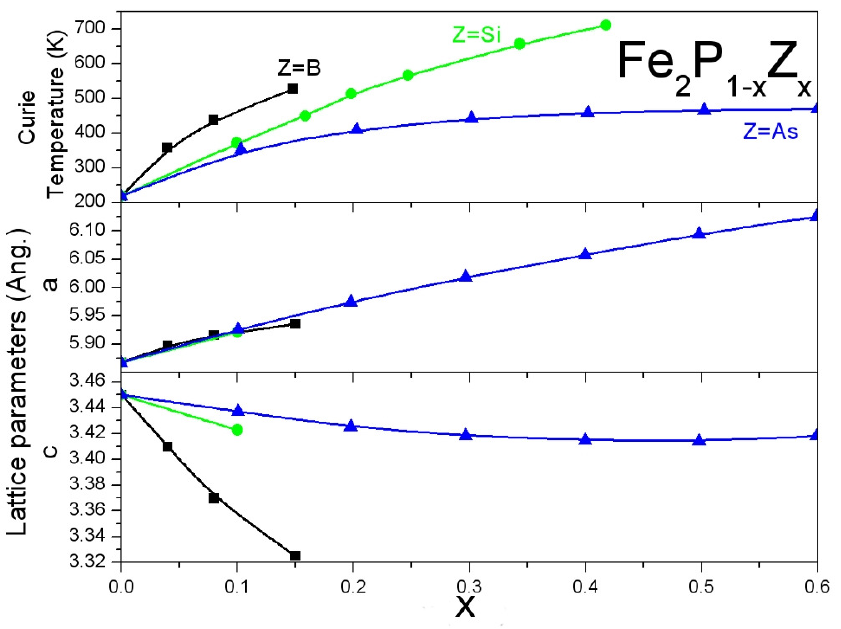}

\caption{\label{fig:doping,-Tc} (Color online) The effect of boron, silicon
and arsenic doping in Fe$_{2}$P$_{1-x}$Z$_{x}$ on the magnetic
ordering temperature $T_{C}$ (top) and the room temperature lattice
parameters (bottom).\cite{Fe2Pdoped1,Fe2Pdoped2,Fe2Pdoped3} }
\end{figure}

\section{Previous experimental findings\label{sec:Previous-experimental-findings}}

\begin{figure}
\includegraphics[width=8.5cm]{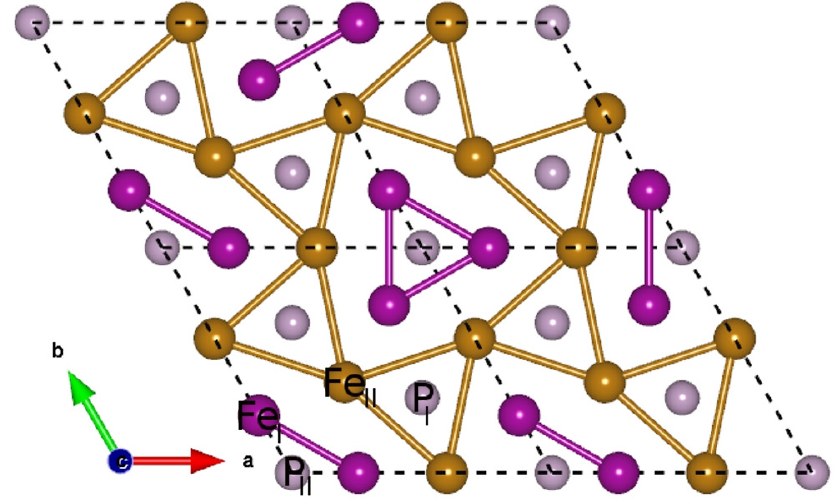}\caption{\label{fig:Structure_Fe2PI} (Color online) Atomic arrangement of
the Fe$_{2}$P in the basal plane. Fe atoms occupy two non-equivalent
three-fold symmetry sites, 3f (Fe$_{I}$) and 3g (Fe$_{II}$), and
the phosphorus atoms sit on a single-fold 1b (P$_{I}$) and two-fold
position, 2c (P$_{II}$) in the hexagonal crystal lattice. Fe$_{I}$
($x{}_{I}$, 0, 0) atoms share the same plane with P$_{I}$ (1/3,
2/3, 0) and Fe$_{II}$ atoms ($x{}_{II}$, 0, 1/2) with P$_{II}$
(0,0, 1/2), respectively.}
\end{figure}

Fe$_{2}$P is the prototype structure of the hexagonal space group
189 ($P\overline{6}2m$) with 9 atoms in the unit cell. The 6 Fe atoms
occupy two non-equivalent threefold symmetry sites (3f and 3g), while
the phosphor atoms sit on a singlefold (1b) and on a twofold position
(2c) in the crystal lattice. Here we adopt the following established
notation: Fe$_{I}$ for the 3f ($x{}_{I}$, 0, 0) positions, Fe$_{II}$
for the 3g ($x{}_{II}$, 0, 1/2) positions, and P$_{I}$ and P$_{II}$
for the 2c (1/3, 2/3, 0) and 1b (0,0, 1/2) positions of the P atoms,
respectively. The hexagonal cell is composed of triangles in the $ab$
plane as shown in Figure \ref{fig:Structure_Fe2PI}. The iron atoms
that alternate along the $c$-axis are surrounded either by four P-atoms
with tetrahedral symmetry (Fe$_{I}$) or by five P-atoms forming a
pyramid (Fe$_{II}$). It was reported previously\cite{FujiiFe2Plattice}
that the $c$-axis lattice parameter increases with temperature while
the $a$- and $b$-axes (basal plane) exhibit negative thermal expansion.
The same authors used a strain gauge dilatometer to measure the linear
thermal expansion ($\Delta l/l$) of a single crystal of Fe$_{2}$P
upon cooling through the first order type magnetic transition and
found a sharp increase of $\frac{\Delta a}{a}=0.74\times10{}^{-3}$
with the simultaneous decrease of the $c$-axis: ($\frac{\Delta c}{c}=-0.84\times10{}^{-3}$).

The coupling between the structure and magnetism is further manifested
by the effect of doping shown in Fig.~\ref{fig:doping,-Tc}. A significant
increase in Curie temperature upon partial replacement of the P atoms
by other $p$-block elements such as B, Si and As has been observed.~\cite{Fe2Pdoped1,Fe2Pdoped2,Fe2Pdoped3}
Catalano et al.~\cite{Fe2Pdoped1} investigated the effect of isovalent
substitution of As for P on the crystalline lattice and on magnetic
properties. The authors found that the larger atomic radius As can
be continuously accommodated into the hexagonal lattice between $0\leq x\leq0.65$
in Fe$_{2}$P$_{1-x}$As$_{x}$ with a strong site preference for
the P$_{I}$ (2c) position. The monotonic increase of the lattice
volume with As addition is a compromise of a moderate increase of
$a$-axis and a simultaneous contraction of the lattice along the
$c$-direction. The same tendency was observed by the substitution
of larger atomic radius, non-isovalent Si for P by Jernberg et al.~\cite{Fe2Pdoped3}
Interestingly, Si atoms were also found to show some site preference
for the P$_{I}$ position with opposing trends in the $a$ and $c$
lattice parameters and a significant increase in the magnetic ordering
temperature. Silicon addition above $x\leq0.1$ results in a change
from the hexagonal lattice structure into one with a body-centered
orthorhombic, \emph{Imm2} (44) symmetry.

Finally, the partial replacement of P by the substantially smaller
atomic radius boron results in a  remarkable increase of the $T_{C}$.
The solid solubility of the B atoms in Fe$_{2}$P$_{1-x}$B$_{x}$
is, however, limited to $x\approx0.15$ due to the formation of other
Fe$_{5}$PB$_{2}$, Fe$_{3}$B and Fe$_{2}$B refractory borides at
higher B concentrations. Chandra et al. established by means of Mössbauer
spectroscopy that the small boron atoms, unlike the larger elements
(As and Si), occupy the P$_{II}$ (1b) singlefold position in the
hexagonal lattice.\cite{Fe2Pdoped2} Another consequence of the chemical
pressure on the hexagonal lattice caused by boron addition as compared
to the parent alloy or to the Si/As-doped compositions is decreased
lattice volume. Despite such differences, the monotonic increase of
the $a$ and decrease of the $c$ lattice parameters of the unit cell
is strikingly similar to the effect of larger $p$-element dopants
(Fig. \ref{fig:doping,-Tc}). An additional consequence of boron doping
is that the first-order PM-FM transition of undoped Fe$_{2}$P becomes
second order.

\section{Methods}

\subsection{Experimental Methods}

The samples used in this study were prepared from ultra high purity
elements. The powders were mixed to weight ratios to according to
nominal compositions of Fe$_{2}$P and Fe$_{2}$P$_{0.96}$Z$_{0.04}$,
(Z=B or C) and then thoroughly ground together in an agate mortar
under protective atmosphere. The initial powders were then pressed
into pellets and sealed into quartz ampoules under protective argon
atmosphere for solid state reaction. The initial annealing temperature
was raised slowly (0.5~K min$^{-1}$) up to 673~K, where each sample
was kept for 4 hours for an initial reaction. There was a subsequent
heating step (at 1~K min$^{-1}$) up to 1273~K where the temperature
was held for a further 4 hours. An additional heat treatment at 973~K
for 4 hours was applied before the sample was oven-cooled to room
temperature. X-ray diffractometry  was used to evaluate the structural
properties of the prepared samples. Single phase hexagonal compositions
were found in all specimens by this method. Neutron diffraction was
carried out at the time-of-flight high resolution powder diffractometer
 at ISIS, UK. This instrument has a resolution of $\frac{\Delta d}{d}$
of 10$^{-4}$ and was used at temperatures between 4.2 and 550~K.
Neutron diffraction found some traces of unreacted carbon in Fe$_{2}$P$_{0.96}$C$_{0.04}$
which we here refer to as a nominal composition since the actual carbon
concentration of the main phase will be somewhat smaller. Magnetic
properties were measured between 10 and 400~K in a Quantum Design
Physical Properties Measurement System .

\subsection{Computational model \label{sub:Theoretical-model}}

The electronic structure calculations were performed using the Vienna
ab initio simulation package (VASP) code, based on DFT within projector
augmented wave  method\cite{VASP} with Perdew-Burke-Ernzerhof  parameterization.\cite{PBE}
Site-based magnetic moments were calculated using the Vosko-Wilk-Nusair
interpolation \cite{Vosko} within the general gradient approximation
 for the exchange-correlation potential. A $k$-point grid of 11 $\times$
11 $\times$ 13 was used to discretize the first Brillouin zone and
the energy convergence criterion was set $10^{-7}$~eV during the
energy minimization process. The density of states (DOS) plots presented
in this work were calculated on a dense (19 $\times$ 19 $\times$
21) $k$-grid for high accuracy. The spin-orbit interaction was turned
off during the calculations and only collinear FM and non-magnetic
(NM) configurations were considered.

The minimal, nine-atom basis cell (six Fe atoms and three P atoms)
was used to evaluate the total energies and magnetic properties of
the alloys. Using this simple model, the effect of doping was simulated
by the replacement of a single phosphorous atom by another $p$-block
element (Z) that represents an $x$=1/3 compositional change in the
Fe$_{2}$P$_{1-x}$Z$_{x}$ formula. Although this approach is undoubtedly
oversimplified in respect of exact compositions provided in the experimental
section, we believe it is still a suitable model to capture the relevant
changes in the electronic structure caused by the dopant elements.
In order to be consistent with the experimental results, we only considered
changes along the $a$- and $c$-axis by the individual expansion
and compression of the $a$- and $c$-lattice parameters, without
allowing any relaxation of the strained structure. In practice, we
varied the lattice parameters using a stepsize of $\pm$0.5\%, calculating
the self-consistent electronic structures at each step.

\section{Results}

\label{sec:Results}

\subsection{Magnetometry}

\begin{figure}
\includegraphics[width=8.5cm]{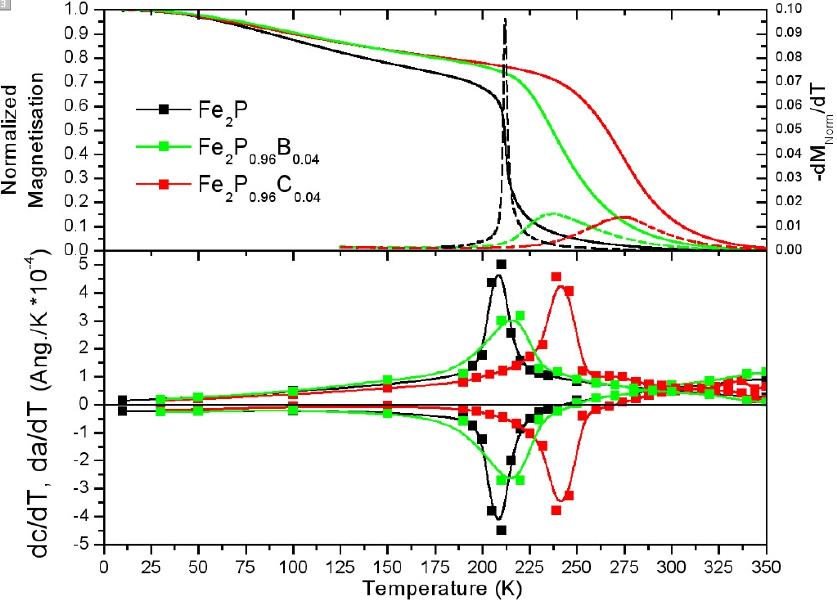}

\caption{\label{fig:magnetisation} (Color online) Thermomagnetic curves of
the parent Fe$_{2}$P alloy with the C and B doped counterparts (top,
left axis) in the field of $\mu_{0}M$=0.01T. Derivatives ($\frac{dM}{dH}$)
are linked to the right axis (top). Bottom figure shows the change
of the lattice parameters ($a,c$) for comparision, obtained using
HRPD.}
\end{figure}

The samples were first cooled to 10~K, where magnetisation curves
were collected at fields up to 5~T, after which thermo-magnetic curves
were collected on heating to 400~K in an applied field of 0.01~T.
The results are plotted in Fig.~\ref{fig:magnetisation}. The parent
alloy shows a sharp, first order transition at $T\approx$215~K in
accordance with values reported previously. The strong effect of doping
on the magnetic order temperature is apparent. As expected, the partial
replacement of P by the much smaller atomic radius B significantly
increases $T_{C}$. In this study, we also partially replaced P by
C atoms for the first time. The empirical atomic radius of carbon
is much smaller (70~pm) than that of the phosphorus (100~pm) and
boron (85~pm) elements. On the other hand, in terms of valence electrons
the sequence B($p{}^{1})<$C$(p{}^{2})<$P$(p{}^{3}$) stands. In
practice, carbon doping shows a very similar effect to that of the
other $p$-block substituents as it also clearly increases the magnetic
order temperature. Furthermore, the Curie transition is heavily broadened
by doping as demonstrated by the smearing out of the temperature derivative
of the magnetisation ($\frac{\partial M}{\partial H}$) (Fig.~\ref{fig:magnetisation}).
Finally, the saturation magnetisation ($M{}_{S}$) increases slightly
with doping; $M{}_{S}$=112 , 113 and 116 Am$^{2}$/kg was obtained
for Fe$_{2}$P, Fe$_{2}$P$_{0.96}$C$_{0.04}$ and Fe$_{2}$P$_{0.96}$B$_{0.04}$,
respectively at 10 K in 9 T applied magnetic field. This slight increase
of $M{}_{S}$ is in line with the expectations from DFT calculations
(see sec. \ref{sub:The-Effect-of-Doping}). However, the samples were
not fully saturated due to the large magnetocrystalline anisotropy.\cite{Fujii}

\subsection{High resolution neutron diffraction}

\begin{figure}
\includegraphics[width=8.5cm]{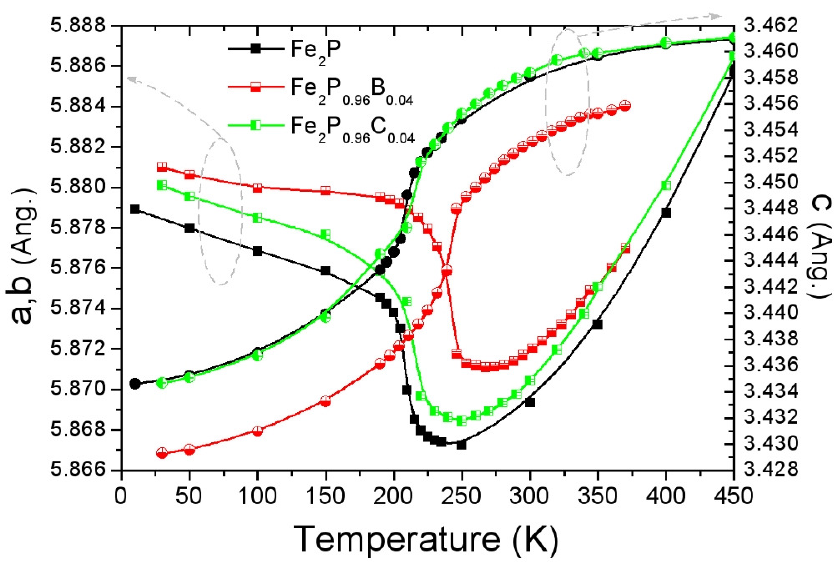}

\caption{\label{fig:a-c_neutron} (Color online) Temperature evolution of the
lattice parameters of the parent Fe$_{2}$P compound together with
that of the C- and B-doped samples. The strong magnetoelastic response
is especially apparent around the magnetic ordering temperature ($\sim$215~K).}
\end{figure}

Fig.~\ref{fig:a-c_neutron} shows the anisotropic lattice expansion
of the $a$- and $c$-axes of Fe$_{2}$P and the doped compounds.
In the magnetically ordered state ($T\lesssim$200~K) the basal plane
exhibits negative lattice expansion with increasing temperature. When
the temperature reaches $T_{C}$, a sharp contraction of the lattice
in the $ab$ plane is observed. The $a$-lattice expansion only looks
Debye-like at higher temperatures ($T\gtrsim$240~K) in the paramagnetic
phase. On the other hand, the thermal expansion of the $c$-axis is
found to be positive over the entire investigated temperature range.
The magnetic ordering temperature is also strongly reflected in the
lattice response along $c$. The consequence of these counteracting
lattice parameter changes over the magnetic transition is a volume
change at $T_{C}$ which is as small as 0.01\%.

\begin{figure}
\includegraphics[width=8.5cm]{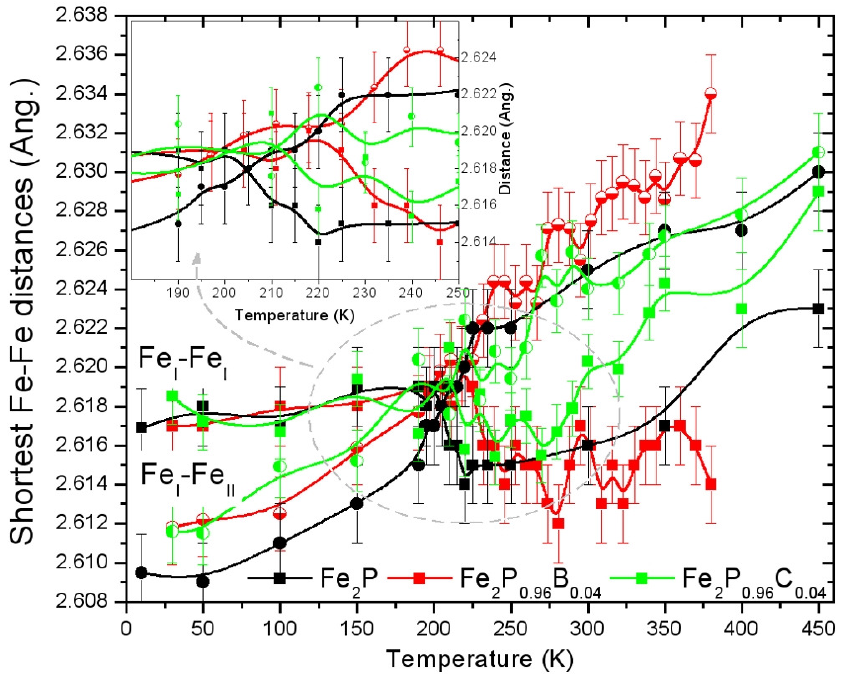}

\caption{\label{fig:a-c_neutron Fe-Fe} (Color online) The shortest Fe$_{I}$-2Fe$_{I}$
and Fe$_{I}$-2Fe$_{II}$ distances as a function of temperature for
all investigated samples. Inset depicts the temperature region of
the crossover.}
\end{figure}

The effect of doping on the lattice parameters is clear; the basal
plane of the hexagonal lattice expands while the c-axis shows contraction.
At first glance, the anomalous thermal lattice expansion resembles
that of the parent alloy. However, temperature derivatives of these
quantities ($\partial a/\partial T$, $\partial c/\partial T$) reveal
characteristic differences between the parent and doped alloys, shown
in Fig. \ref{fig:magnetisation}. The sharpness of the derivatives
demonstrates the first order nature of the magnetoelastic transition
of stoichiometric Fe$_{2}$P. Although $\partial a/\partial T$ and
$\partial c/\partial T$ are of opposite sign, they both have sharp
peaks at temperature $T_{p}$=209~K. Both B and C doping cause a
shift of $T_{p}$ to higher temperatures with broader $\partial a/\partial T$
and $\partial c/\partial T$. These effects on the lattice properties
correspond very well to the observed changes in the magnetic properties.
The temperature derivative of the magnetisation in Fig.~\ref{fig:magnetisation}
(top), reveals similar broadening of the ferromagnetic transition
with B or C doping.

HRPD is capable of tracking the change in the interatomic distances,
thus providing  information about the magnetoelastic coupling in these
materials. The evolution of the lattice parameters through the magnetoelastic
transition is already a clear indication of the significant changes
of the atomic distances. Both the 3f and 3g positions of the iron
atoms are low symmetry positions described by the positional parameters
$x{}_{I}$ and $x{}_{II}$, respectively. Iron atoms on the 3f-site
(Fe$_{I}$) are connected to 2 other iron atoms of the same type,
denoted as Fe$_{I}$-2Fe$_{I}$. There are also two distinctive Fe$_{I}$-2Fe$_{II}$
and Fe$_{I}$-4Fe$_{II}$ connections to iron atoms located at the
3g-site (Fe$_{II}$). Finally, there exists a relatively close Fe$_{II}$-4Fe$_{II}$
distance above 3~\AA{}. In Fig. \ref{fig:a-c_neutron Fe-Fe}, we
only plot the two shortest distances, Fe$_{I}$-2Fe$_{I}$ and Fe$_{I}$-2Fe$_{II}$,
as a function of temperature, evaluated from Rietveld refinements
of the HRPD data. In all of the investigated samples, the Fe$_{I}$-2Fe$_{II}$
distance is the shortest Fe-Fe separation at low temperatures($<$200~K).
The latter increases with temperature, eventually becoming larger
than the Fe$_{I}$-Fe$_{I}$ distance, which is strongly reduced in
the vicinity of the magnetic transition temperature. The point in
temperature above which Fe$_{I}$-Fe$_{I}$ is the shortest Fe-Fe
distance is between 190 and 220~K for all samples.

\begin{figure}
\includegraphics[width=8.5cm]{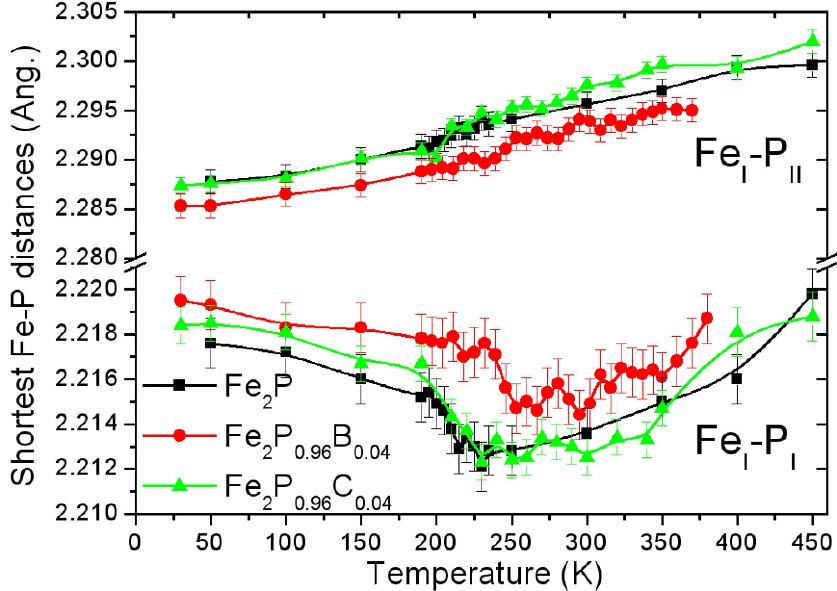}

\caption{\label{fig:a-c_neutron-Fe-P} (Color online) The shortest metal-metalloid
distances, Fe$_{I}$- 2P$_{I}$ and Fe$_{I}$- 2P$_{II}$ as a function
of temperature in Fe$_{2}$P together with the doped counterparts.
Both distances relate to the metamagnetic 3f (Fe$_{I}$) site.}
\end{figure}

In addition we can distinguish two groups of metal-metalloid distances:
Fe$_{I}$-(P$_{I}$,P$_{II}$) and Fe$_{II}$-(P$_{I}$,P$_{II}$).
The shortest one is Fe$_{I}$-2P$_{I}$, followed by Fe$_{I}$-2P$_{II}$,
Fe$_{II}$-P$_{II}$ and Fe$_{II}$-4P$_{I}$. Fig. \ref{fig:a-c_neutron-Fe-P}
contains the two shortest distances only, for the first time. The
separation of Fe$_{II}$-P$_{II}$ and Fe$_{II}$-4P$_{I}$ atoms
is in the range of $\sim$2.37~\AA{} and $\sim$2.49~\AA{}, respectively.
It is worth noting that both Fe$_{I}$-2P$_{I}$ and Fe$_{II}$-P$_{II}$
(not shown)  only have components within the $ab$-plane which explains
the resemblance of their temperature evolution to that of the $a$
lattice parameter in Fig.~\ref{fig:a-c_neutron}: the shortest Fe$_{I}$-2P$_{I}$
distance decreases sharply on heating through  the magnetic transition
 in the parent alloy as well as  in the doped compounds. Its significance
will be discussed in the context of our theoretical results in the
next section. The larger $a$ lattice parameter in the doped compounds
means that the  Fe-P distance is largest in those samples. 

\section{Theoretical results \label{sec:results-of-theoretical}}

\subsection{Stoicheometric Fe$_{2}$P}

\begin{figure}
\includegraphics[width=8.5cm]{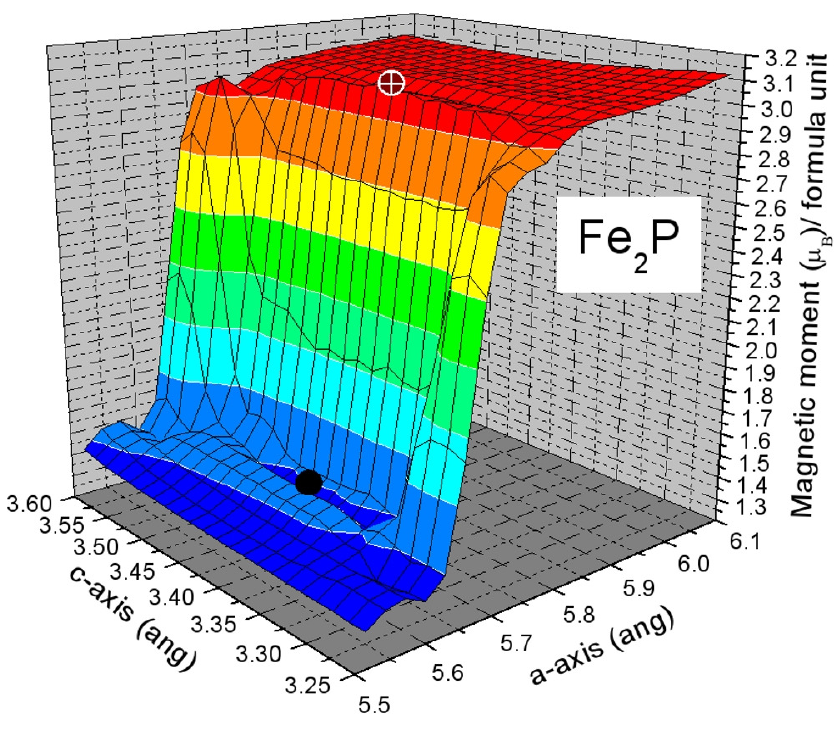}

\caption{\label{fig:Fe2P_Mtot}Total magnetic moment / formula unit as a function
of lattice parameter of Fe$_{2}$P. The magnetic moment  is strongly
linked to the change in \emph{a}-lattice parameter. The red and black
dots indicate the high and low magnetisation states (see text).}
\end{figure}

\begin{figure}
\includegraphics[width=8.5cm]{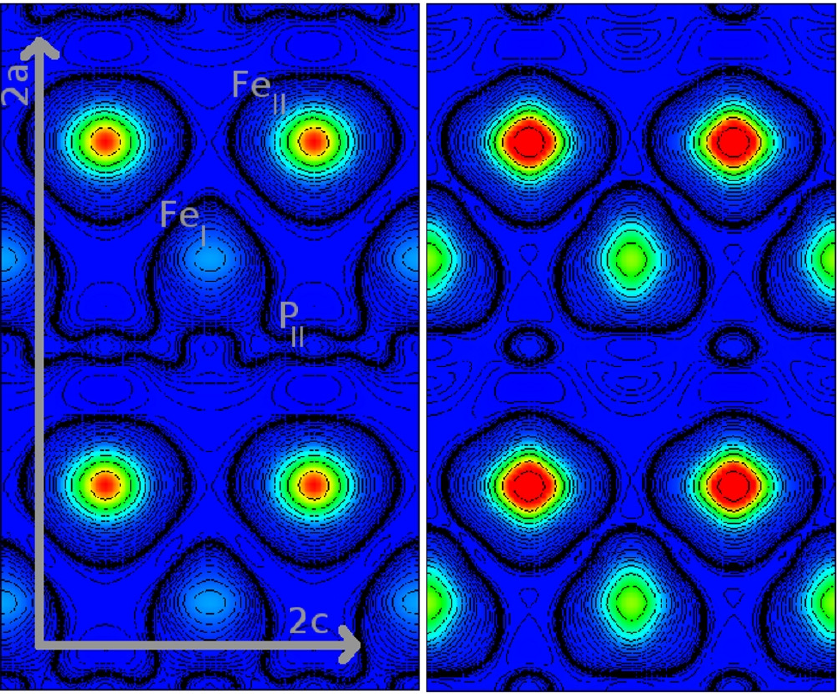}

\caption{\label{fig:ac_spindensity} (Color online) A comparison of the spin
density of the low magnetisation state (left) and high magnetisation
state (right) and of Fe$_{2}$P. The large magnetic moment around
the Fe$_{II}$ site is only decreased in amplitude above the transition,
when strong delocalisation of the Fe$_{I}$ magnetisation occurs simultaneously.}
\end{figure}

We now use a zero temperature theoretical model to explore the effect
of changing the lattice structure on the magnetisation of Fe$_{2}$P
in this subsection and of the doped compounds in the following section.
Our aim is to establish the most significant contributions to the
magneto-elastic coupling in this material family. The calculated variation
of total magnetisation in the FM state with lattice parameters  is
plotted in Figure \ref{fig:Fe2P_Mtot}. A small change in the lattice
can have a very strong effect on the magnetic properties. The lattice
expansion within the basal plane causes the total magnetic moment
/ formula unit to increase abruptly from $\sim1.3\mu_{B}$ to 3.1~$\mu_{B}$
around a critical value of $a\approx5.7$~\AA{}, and stays practically
constant above it. The magnetisation varies very little with $c$,
reflecting the greater importance of the atomic distances within the
basal plane, where the atoms are packed more densely. The calculations
reveal the unusual duality of the magnetic structure of Fe$_{2}$P
in which there is a large magnetic moment of $M$(Fe$_{II}$)=2.16~$\mu_{B}$
on the 3g-site together that coexists with a significantly smaller,
$M$(Fe$_{I}$)=0.85~$\mu_{B}$, moment on the 3f crystallographic
site. These findings are in full agreement with previous studies (see
Sec.~\ref{sec:Previous-experimental-findings}). Fig.~\ref{fig:ac_spindensity}
compares the differences in spin densities of the high and low magnetic
configurations (as indicated by the red and black dots in Fig. \ref{fig:Fe2P_Mtot}).
As a consequence of the lattice symmetry, the shortest metal-metalloid
distance, Fe$_{I}$-P$_{II}$, as well as the shortest metal-metal
separation, Fe$_{I}$-Fe$_{II}$, have projections along both the
$a$- and $c$-axes, and so this is the most suitable cross-section
for our analysis. If we consider solely the effect of the structural
changes that we know to occur at the Curie transition (Fig.~\ref{fig:a-c_neutron Fe-Fe}),
we find that there is a redistribution of magnetisation of the Fe$_{I}$-site.
The large and localised magnetic moment on the Fe$_{II}$ site is
decreased in amplitude to $\sim1.3\mu_{B}$ above the transition and
strong delocalisation of the Fe$_{I}$ magnetisation occurs simultaneously.
Electrons from the Fe$_{I}$ site that have a magnetic contribution
are redistributed along the $c$-axis in directions that link the
Fe$_{I}$ and P$_{II}$ sites.

\begin{figure}
\includegraphics[width=8.5cm]{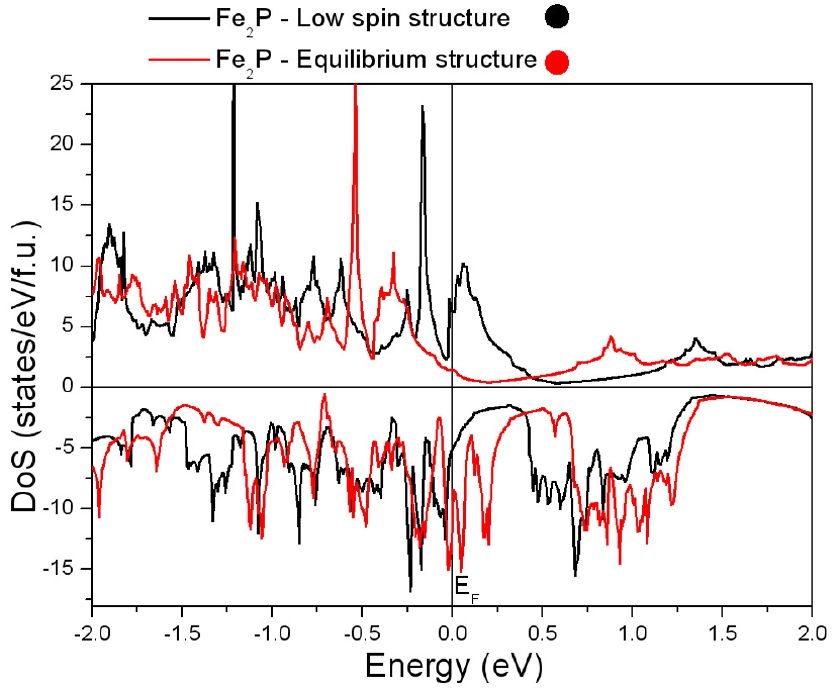}

\caption{\label{fig:Fe2P_DOS-1} (Color online) Comparision of the density
of states of a high magnetisation state (red) and a low magnetisation
state (black) in Fe$_{2}$P. }
\end{figure}

 In Fig. \ref{fig:Fe2P_DOS-1}, we compare the total electronic density
of states  of the high and low magnetisation states, at ($a,c$) values
indicated by the red and black dots in Figure~\ref{fig:Fe2P_Mtot}.
 The high magnetisation state has a high and sharp DOS at $E_{F}$
with $\sim$80\% spin polarisation, dominated by minority (down) spins
(Figure~~\ref{fig:Fe2P_DOS-1}, red line).  When the lattice is
compressed within the basal plane, the magnetisation  is significantly
weakened, as shown  in Figure~\ref{fig:Fe2P_Mtot}. The filled bands
at low energy  contain the phosphorus 3$s$ states (not shown). The
conduction band is formed by the 3$d$ states of Fe and the 3$p$
states of P as is typical of strong $p$-$d$ hybridization. The hybridized
states are also present near the Fermi level, influencing the exchange
splitting of the 3$d$ states of Fe. The effect of the smaller $a$-lattice
parameter (in the low magnetisation state) is to reduce the exchange
splitting considerably. The high peaks in the majority (up spin) DOS
located around -0.5~eV are shifted to -0.2~eV. In the minority DOS,
the opposite trend is observed as the triple peak feature between
0 and 0.25~eV is lowered to around -0.25~eV.

We can ascribe three competing effects to the magnetoelastic transition.
From our experiment, with increasing temperature the separation between
Fe$_{I}$ and Fe$_{II}$ atoms increases monotonicly  was observed
(see Fig. \ref{fig:a-c_neutron Fe-Fe}).  Our previous exchange coupling
analysis~\cite{Erna_exchange} found that the largest contribution
to the decrease in the $c/a$ ratio  is the weakened magnetic interaction
between the Fe atoms on 3f and 3g sites. Secondly, at the magnetic
transition the Fe$_{I}$-Fe$_{I}$ distance is also altered and it
becomes comparable to Fe$_{I}$ - Fe$_{II}$ but the exchange constant
between Fe$_{I}$-Fe$_{I}$ atoms is only the fraction of the Fe$_{I}$-Fe$_{II}$
exchange, and has a less pronounced dependence on distance. Thirdly,
the shortest Fe$_{I}$-P$_{I}$ distance, located in the $ab$ plane
also  decreases around the transition temperature. This results in
a significantly stronger hybridisation that can explain the decrease
in magnetisation and alternation of these bonds, and hence trigger
a lattice distortion in the first place. A similar scenario was also
drawn from the comparison of the total electronic charge density in
the FM and PM states of Mn and Si doped Fe$_{2}$P-type alloy by Dung
et al.~\cite{Dung1} 

In the following section, we compare the effect of doping on the structure
and magnetic properties obtained by means of DFT using structural
data from neutron diffraction.

\subsection{The Effect of Doping \label{sub:The-Effect-of-Doping}}

\begin{figure}
\includegraphics[width=8.5cm]{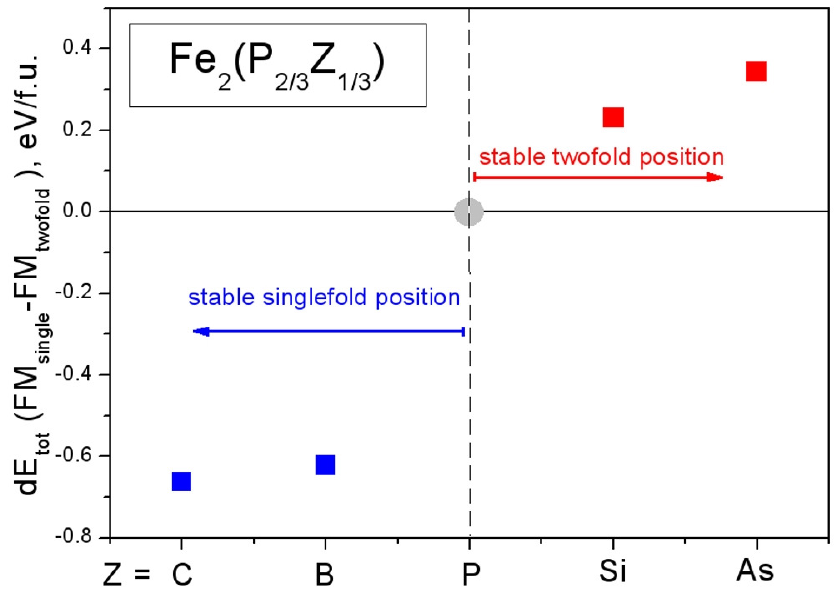}

\caption{\label{fig:sitepreference} (Color online) Site preference of the
$p$-block dopants as established from DFT. Negative total energy
differences represent preference for the single-fold position. Positive
total energy differences suggest tendency for two-fold site occupancy. }
\end{figure}

There exist two crystallographic sites where the metalloid elements
can be incorporated into the lattice, albeit with strong site preference
as described in Sec.~\ref{sec:Previous-experimental-findings}. Experimentally,
elements with an atomic radius smaller than P were found to show a
preference in the single-fold position (1b), whilst larger elements
tends to occupy the twofold symmetry site (2c) in the hexagonal lattice.
In order to establish the site preference from DFT calculations, we
compare the difference in total energy for the different elemental
substitution. First of all, the calculated total energies of the FM
solutions (at any $(a,c)$ lattice parameter value) were always found
to be energetically more stable than that of the NM configurations
by  1.5-3~eV/f.u., yielding a strong tendency toward the formation
of magnetic order in all the investigated alloys. The site preference
of the dopants can be established by calculating the difference in
total energy ($dE$) between the single-fold and two-fold site position
occupancies using the equilibrium lattice of the FM state. The calculations
find a clear trend with the size of the substituents: the energy difference
is negative for elements that are smaller than P (left hand side of
Fig.~\ref{fig:sitepreference}), reflecting the preference for singlefold
occupation. On the other hand, a substituent with larger atomic radii
(Si and As) prefers to occupy the twofold position of the hexagonal
lattice.

This theoretically established site occupation of the dopant elements
is in line with the experimental findings see Sec.~\ref{sec:Previous-experimental-findings},
suggesting the dominance of size effects. Therefore, we hereafter
focus our magnetoelastic investigations on the effect of single-fold
site occupation by the small C and B elements and the two-fold site
occupation of the large Si and As elements.

\begin{figure}
\includegraphics[width=8.5cm]{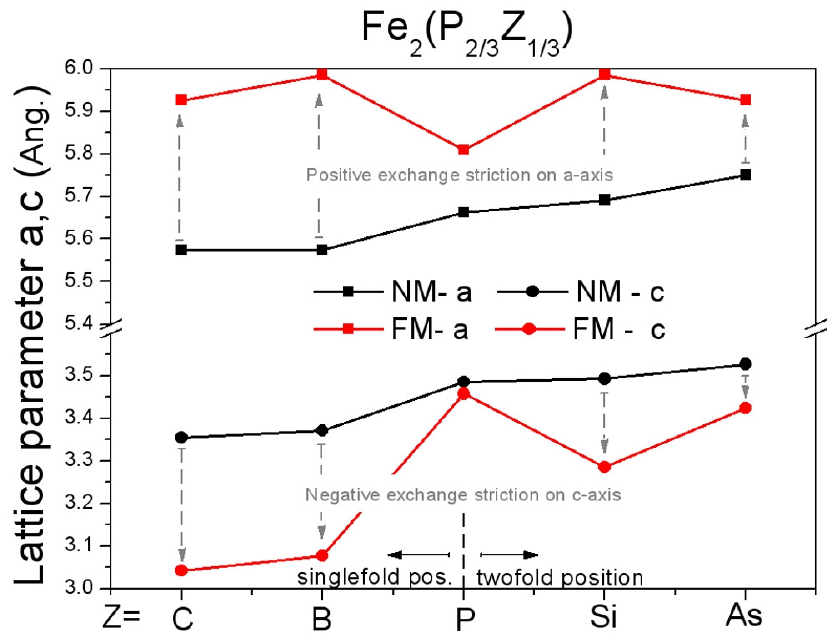}

\caption{\label{fig:Exchange_striction} (Color online) Equilibrium lattice
parameters for NM and FM states as a result of DFT calculations as
a function of dopants in Fe$_{2}$P$_{2/3}$Z$_{1/3}$, where Z=C,
B, Si, or As, respectively.}
\end{figure}

The effect of doping on the lattice parameters is shown in Fig.~\ref{fig:Exchange_striction}.
The NM calculations (black line) show an increasing trend in both
the $a$- and $c$ parameters of the hexagonal structure with the
inclusion of larger atomic radius, anionic $p$-block elements. This
behaviour is easily anticipated in terms of chemical pressure as the
lattice is expected to adapt according to the size of dopant. Examining
Fig. \ref{fig:Exchange_striction}, the steepest change in the lattice
is seen with the partial replacement of P by B atoms, due to the largest
difference in atom size between these elements. However, in the presence
of FM interactions, any monotonic relation of lattice parameter to
the atomic radius of the dopant is broken. The parent alloy has significantly
smaller $a$ and larger $c$ parameters in the FM ground state than
the doped compounds. The onset of ferromagnetism increases the separation
of atoms within the basal plane (positive exchange striction) while
their separation along the $c$-axis is reduced (negative exchange
striction), regardless of the size or valence state of the dopant
as compared to the parent composition (Fe$_{2}$P). The overall volume
change caused by the ferromagnetic exchange is positive, dominated
by the larger $a$ parameter ($V_{hex}=a^{2}c\sin(2\pi/3)$). Based
on these calculations, one would expect the lattice to expand along
the $a$-axis with a simultaneous contraction of the $c$-axis in
the vicinity of the magnetic ordering temperature. Indeed, experimental
(HRPD) measurements clearly reveal this sharp positive exchange striction
in the basal plane and the shrinkage of the $c$-axis upon cooling
the sample through the Curie temperature (see Fig.~\ref{fig:a-c_neutron}).
\begin{figure}
\includegraphics[width=8.5cm]{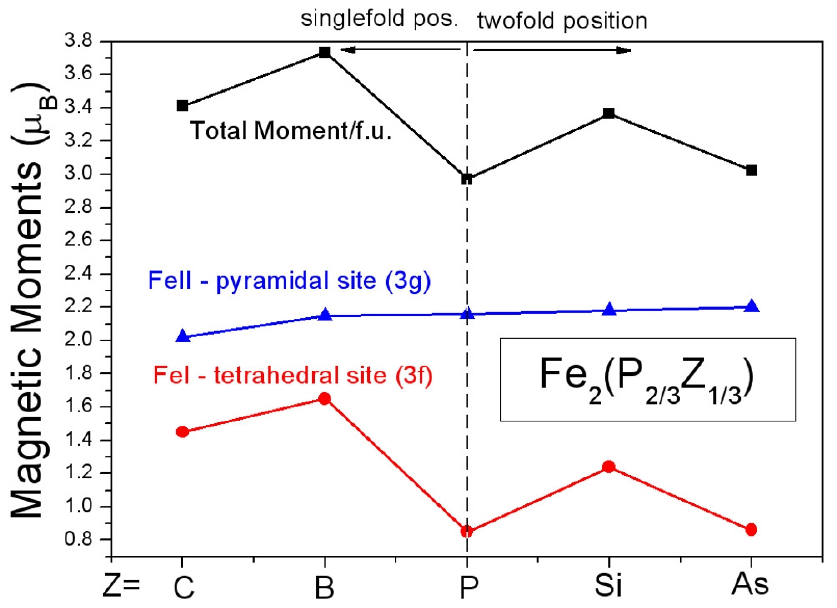}

\caption{\label{fig:M_eq} (Color online) Total and site-projected magnetic
moments in Fe$_{2}$P$_{2/3}$Z$_{1/3}$, for Z=C,B,Si and As, respectively. }
\end{figure}

The significant coupling of the lattice  to the magnetisation also
becomes aparent from the analysis of the total and site-projected
magnetic moments as plotted in Fig. \ref{fig:M_eq}. The magnetic
moment of Fe$_{II}$ atoms show very little variation with doping
and stays at around 2.2~$\mu_{B}$. Although the larger $p$-block
dopants increase the separation of the Fe$_{II}$-Fe$_{II}$ atoms,
this distance is typically above 3~\AA{}, and so doping only slightly
alters the exchange-split states. In strong contrast, the site-resolved
magnetisation of the 3f site is found to be much more sensitive to
the valence electron number of the anionic elements. Examining Fig.~\ref{fig:M_eq}
there is a strong relation between added valence electron number and
magnetisation with Z=B ($-2e{}^{-}$), C ($-1e{}^{-}$), P($-$),
Si ($-2e{}^{-}$), or As($-$). This feature can be explained in terms
of charge transfer from Fe$_{I}$ to the anionic Z element in order
to fill the electronegative $p$ shell of the latter. The charge transfer
is distance dependent and is strongly supressed by the strong $p$-$d$
interaction at smaller Fe$_{I}$-Z separations, while lattice expansion
suppresses $p$-$d$ hybridization and increases the degree of localisation
and ionic bonding in the system. With doping, the short Fe$_{I}$-Z
distances are strongly influenced, altering the balance between magnetic
moment formation (Fig.~\ref{fig:ac_spindensity}) and bond formation.
The larger magnetic moments on the Fe$_{I}$ atoms that result from
doping account for the expansion of the basal plane as indicated by
the comparison of Figs.~\ref{fig:Exchange_striction} and~\ref{fig:Fe2P_Mtot}.

\section{Summary and Conclusion}

We have used high resolution neutron diffraction  as well as density
functional theory to investigate the effect of $p$-block element
doping on the magnetoelastic properties of Fe$_{2}$P. HRPD has revealed
that the strong coupling between the magnetism and the lattice is
manifested by the contraction of the basal plane and by a significant
increase of the $c$-axis on heating through the magnetic transition,
resulting in a small overall volume change of the lattice ($>$0.01\%).
A simultaneous change in both the metal-metal and metal-metalloid
distances is observed.

DFT calculations reveal that the magnetic properties are strongly
dependent on expansion in the basal ($ab$) plane, while they are
almost invariant with regard to variation of the $c$ parameter. The
strong dependence of magnetic moments on the 3f site is related to
the $a$ lattice parameterthat lies in the proximity  of a critical
value where change from low to high magnetisation can occur.  The
closest metal-metal and metal-metalloid distances - the latter  also
linked to the metamagnetic 3f site - are strongly altered by both
the $d$-$d$ and $p$-$d$ hybridisation energies at the transition.
As a result, the redistribution (delocalisation) of the magnetisation
from the 3f site along the Fe$_{I}$-P$_{II}$ chains in the $c$-axis
direction occurs as shown in Fig. \ref{fig:ac_spindensity}, implying
its strong influence on bonding (also suggested by Dung et al.\cite{Dung2})
Our current analysis finds the survival of the magnetic moment on
the 3g site above the magnetic transition. This is in line with our
previous exchange interaction analysis \cite{Erna_exchange} which
revealed anisotropic coupling, with strongly ferromagnetic Fe$_{II}$-Fe$_{II}$
interactions accounting for the short range magnetic correlations
and the observed divergence of Curie-Weiss law.\cite{Fujii,Dung2}

Our theoretical investigations on the effect of doping show that elements
with an atomic radius smaller than phosphorus (e.g. C or B) occupy
single-fold (1b) sites while larger elements such as Si or As prefer
to occupy the two-fold symmetry site (2c) in the hexagonal lattice.
The moment on the   3g site  is not influenced by the dopant and takes
a high value  ($\sim$2.2$\mu_{B}$). By contrast, the iron atoms
on the metamagnetic site develop smaller  magnetic moments, the magnitude
of which varies in direct relation to the valence electron number
of the doping element. As a result, the conditions for the metamagnetic
transition of the 3f site are locally modified by doping, resulting
in a smearing of the transition as observed experimentally. Indeed,
the increased  exchange splitting on Fe$_{I}$ results in increased
exchange coupling parameters, significantly increasing the Curie temperature.\cite{Erna_exchange}

Our DFT analysis finds that the magnetisation is highly sensitive
to the change in lattice dimensions,  therefore both compositional
inhomogeneities and internal strains as set by the preparation conditions
can significantly alter the observed magnetic properties of the material.\cite{priv:Gercsi_private_com}
. For applications (such as magnetic refrigeration), where the transition
temperature needs to be set by the material to a high accuracy, elements
such as Si and As are favorable but process control will be crucial.
Finally, as the remarkable change in magnetic ordering of these alloys
is linked to the lattice parameters within the basal plane, uniaxial
thin films on flexible substrates could be exploited to electrically
manipulate the magnetic properties of this material system.
\begin{acknowledgments}
We thank L. Szunyogh for useful discussions and D. Boldrin for help
with sample preparation. We also thank K.S. Knight for providing us
with the $^{11}$B isotope used for this study. Financial support
is acknowledged from The Royal Society (KGS) and EPSRC grant EP/G060940/1
(KGS and ZG). The research leading to these results has received funding
from the European Community's 7th Framework Programme under Grant
agreement 310748 ``DRREAM''. Fig.~\ref{fig:Structure_Fe2PI} and
\ref{fig:ac_spindensity} were prepared using VESTA open-source software.~\cite{Momma}
Computing resources provided by Darwin HPC and Camgrid facilities
at The University of Cambridge and the HPC Service at Imperial College
London are gratefully acknowledged.

\bibliographystyle{/home/joy/Dropbox/Papers_in_preparation/Zsolt_Fe2PX/Fe2P}
\bibliography{Fe2P}
 \end{acknowledgments}

\end{document}